# Defect filtering for thermal expansion induced dislocations in III-V lasers on silicon

Jennifer Selvidge[1], Justin Norman[1], Eamonn T. Hughes[1], Chen Shang[1], Daehwan Jung[2], Aidan A. Taylor[1], MJ Kennedy[3], Robert Herrick[4], John E. Bowers[1,3], and Kunal Mukherjee[1a]

[1]*Materials Department, University of California, Santa Barbara, CA 93106, USA*
[2]*Korea Institute of Science and Technology, Seoul 02792, South Korea*
[3]*Electrical and Computer Engineering Department, University of California, Santa Barbara, CA 93106, USA*
[4]*Intel Corporation, Santa Clara, CA 95054, USA*

Epitaxially integrated III-V semiconductor lasers for silicon photonics have the potential to dramatically transform information networks, but currently, dislocations limit performance and reliability even in defect tolerant InAs quantum dot (QD) based lasers. Despite being below critical thickness, QD layers in these devices contain previously unexplained misfit dislocations, which facilitate non-radiative recombination. We demonstrate here that these misfit dislocations form during post-growth cooldown due to the combined effects of (1) thermal-expansion mismatch between the III-V layers and silicon and (2) precipitate and alloy hardening in the active region. By incorporating an additional sub-critical thickness, indium-alloyed 'misfit dislocation trapping layer', we leverage these mechanical hardening effects to our advantage, successfully displacing 95% of misfit dislocations from the QD layer in model structures. Unlike conventional dislocation mitigation strategies, the trapping layer reduces *neither* the number of threading dislocations *nor* the number of misfit dislocations. It simply shifts the position of misfit dislocations away from the QD layer, reducing the defects' impact on luminescence. In full lasers, adding a misfit dislocation trapping layer both above and below the QD active region displaces misfit dislocations and substantially improves performance: we measure a twofold reduction in lasing threshold currents and a greater than threefold increase in output power. Our results suggest that devices employing both traditional threading dislocation reduction techniques and optimized misfit dislocation trapping layers may finally lead to fully integrated, commercially viable silicon-based photonic integrated circuits.

Silicon-based photonic integrated circuits are poised to dramatically increase data network bandwidth and energy efficiency and enable new paradigms in chip-scale sensing, detection, and ranging. Direct epitaxial integration of III-V semiconductor lasers on silicon promises cost-efficiency and scalability,[1] but, fabricating reliable, high-performance GaAs- or InP-based lasers on silicon is challenging.[2–6] Lattice constant mismatch between the substrate and III-V film generates threading dislocations (TDs) that rise upward through the film.[4] Where they intersect the device active region, they facilitate non-radiative recombination, degrading both performance and reliability.[2,7,8] And while decades of work have reduced TD densities (TDD) to $10^6$–$10^7$ cm$^{-2}$ [9–11] and developed more dislocation-tolerant active materials such as InAs-quantum dots (QDs) in InGaAs quantum wells (QW) (dots in a well or DWELL),[12–18] commercially viable III-V lasers on silicon have yet to be realized.

We recently discovered an additional, key performance-limiting defect present in InAs QD lasers on silicon: unexpected misfit dislocations (MDs) lying along the upper and lower boundaries of the active region, even in record lifetime QD lasers.[19,20] These MDs, like TDs, limit performance and reliability because they too are potent non-radiative recombination centers.[21] Worse still, they may be far more damaging as they have a much larger interaction area with the active region. MDs normally form during growth in layers exceeding the critical thickness for dislocation glide;[22] to prevent this, the active layers in both QW and DWELL lasers are carefully designed to be below critical thickness.[23] Yet, both QW- and QD-based devices contain these MDs.[24] However, there are few reports of this, possibly because these defects can easily go unnoticed: the QDs' strain contrast masks the MDs' strain contrast in cross-sectional transmission electron microscopy (XTEM). Thus, MDs have gone unaddressed in QD systems.

Here, we propose a formation mechanism for these MDs centered on thermal expansion mismatch rather than lattice mismatch and validate it in model structures and full DWELL lasers on silicon. We further demonstrate a novel MD filter that consists of thin alloy-hardened "trapping layers" placed directly above and below the laser active region in the epitaxial stack. These layers displace MDs away from the QDs, rather than removing them entirely, an atypical filtering strategy that we show yields improvements comparable to an order of magnitude reduction in TDD.

We first examine model structures using scanning electron microscopy (SEM)-based cathodoluminescence (CL) spectroscopy to directly observe the effects of MDs on DWELL emission. The trapping-layer-free "baseline" structure (described in Ref. 21) consists of a GaAs film with a single InAs DWELL layer 100 nm below the surface of a GaAs-on-Si template.[21] CL images were collected at room temperature on an Attolight Rosa at 10 kV. The CL map of the wetting-layer emission at 1005 nm (Fig. 1a) shows a network of dark lines and spots, corresponding to MDs and TDs, respectively. The sharp dark lines indicate that MDs lie sufficiently close to the QD layer to substantially lower light emission in their vicinity. The InAs QD ground-state luminescence map collected at 1245 nm (Fig. 1b) has these same dark features, although they appear more diffuse due to inhomogeneous strain.

---

[a] Author to whom correspondence should be addressed: kunalm@ucsb.edu

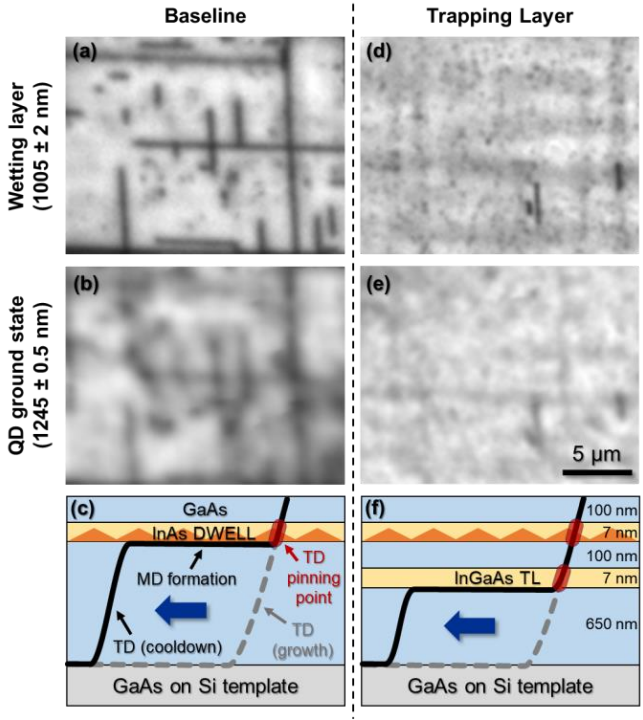

**Fig. 1. (a-b)** Monochromatic cathodoluminescence (CL) images of the baseline structure at **(a)** QD wetting-layer emission wavelength (1005 nm) and **(b)** QD ground-state emission wavelength (1245 nm). **(c)** Schematic representation of approximate dislocation evolution in baseline structure. During cooldown, enough tensile stress builds in the thick GaAs layer below the QD layer for the threading dislocation (TD) to glide from its growth position (gray dotted line). The QD layer pins the TD (red box), causing a misfit dislocation (MD) to form at the bottom interface. The 100-nm GaAs cap is too thin to relax, so no MD forms here. **(d-e)** Comparable CL of the trapping layer structure from **(d)** the wetting layer **(e)** and the QD ground state. Total dark line length in the DWELL layer is 20× lower than in the baseline. **(f)** While the TD in the thick GaAs still glides in response to the tensile stress, by introducing an additional TD pinning point, the trapping layer displaces MD formation away from the QD layer.

The single DWELL layer is below critical thickness, so we hypothesize the MDs in this system form not during growth but during cooldown. Since GaAs has a larger thermal expansion coefficient than silicon ($\alpha_{GaAs} - \alpha_{Si} \approx 3\times10^{-6}$ K$^{-1}$), GaAs layers, which are essentially unstrained by the end of growth at 540 °C, become up to 0.1% biaxially tensile strained during cooldown as they approach 300 °C. Due to this thermal strain, the GaAs layer that exceeds critical thickness with respect to silicon, and the still-mobile TDs can glide in the GaAs layer if thicker than just a few hundred nanometers. This is not surprising: we know thermal stress can drive dislocation glide—thermal cyclic annealing (TCA) takes advantage of this very principle to reduce TDDs.[25] Even so, TDs gliding during cooldown is not inherently problematic. If, however, the DWELL layer pins threading dislocations,[26] as shown in Fig. 1c, then as the free TD segment in the thick GaAs layer glides away, the TD segment in the DWELL is left behind, and a MD will form at the QD layer interface. We hypothesize that the TD pinning force arises from the precipitate-like QDs[27] and from random compositional fluctuations in the In$_{0.15}$Ga$_{0.85}$As QW. This latter effect occurs because the difference in covalent radii of indium (142 pm) and gallium (124 pm) generates in-layer stress fluctuations.[28] This alloy hardening phenomenon has been reported on previously in semiconductors.[26,29,30] See supplementary information (Fig. S1) for additional detail. Finally, note that no TD glide (and thus no MD formation) occurs above the DWELL because the GaAs capping layer is too thin to relax.

If our proposed mechanism is correct, we can leverage this same effect to displace the MDs from the DWELL layer. By inserting a 7-nm In$_{0.15}$Ga$_{0.85}$As "trapping layer" 100 nm below the DWELL, we are able to reduce the MD length at the DWELL layer by 20× (Fig. 1d and 1e). The trapping layer itself should have a negligible impact on the TDD because it is below critical thickness; and, indeed, the measured TDD is similar to the baseline structure. Instead, the layer works by introducing an additional TD pinning point (red box) (Fig. 1f), so TD glide and the resulting MD formation occur only below the trapping layer. We attribute the faint broad dark lines to MDs below the trapping layer. Assuming the distance between the pinning points is sufficiently small (i.e. the intermediate GaAs is below the critical thickness induced by the thermal contraction), no MDs can form between the trapping layer and the DWELL. And, just as with the DWELL, MDs cannot pass through the trapping layer due to the tensile-to-compressive strain reversal at this interface.

To gain more detailed insight into the structural evolution of MDs and TDs, we use a diffraction-based SEM technique, electron-channeling contrast imaging (ECCI), to directly observe a continuation of the MD formation process that occurs during cooldown at room temperature (Fig. 2). Although TD glide ceases below ~300 °C, the thermally induced tensile stress continues to build; the GaAs layers experience a 0.15% biaxial tensile strain at room temperature. Fig. 2a—collected on a Thermo Fisher Apreo SEM at 30 kV in the (400)/(220) channeling condition—shows a time-lapse evolution of a single TD in the baseline model structure. Initially, only a spot of point contrast is visible where the TD segment exits the film surface. Electron-beam irradiation supplies energy that reanimates TD glide, so the unpinned TD segment below the QD-layer pinning point glides away, forming a MD that lengthens over time (orange arrows). We see no point contrast on the growing end indicating that this end sinks down into the film beyond the detection range of ECCI, just as depicted in Fig. 1c. If, instead, MDs formed due to the DWELL exceeding critical thickness, we would expect to see the upper TD segment gliding, but here it is stationary. This provides direct evidence that our proposed mechanism—based on thermal strain buildup during cooldown and local TD pinning—drives MD formation.

Fig. 2b-2e—collected on a FEI Quanta SEM under the same conditions as Fig. 2a—compare MD densities between the baseline and the trapping layer structure before and after heavy electron-beam irradiation. The as-grown baseline structure (Fig.

2b) contains MDs, marked with black arrows, following growth and cooldown. Based on the limited 100-200 nm depth sensitivity of ECCI, these sharp-contrast MDs must be reasonably near the film surface, most likely just below the shallow QD layer, as in Fig. 1a-1c., Electron-beam irradiation causes new sharp-contrast MD segments, marked with orange arrows, to form and grow (Fig. 2c). In the as-grown trapping-layer structure (Fig. 2d), we measure a 20× reduction in total shallow (high-sharpness) MD length from baseline (over a 2500-μm$^2$ area), in agreement with CL. Electron-beam irradiation generates a high-density network of diffuse-contrast MD lines (Fig. 2e). Their diffuse contrast indicates that these dislocations are located deeper in the structure, likely at the trapping layer.[31] Notably, the density of high-sharpness, shallow dislocations remains constant, indicating that SEM irradiation does not increase MD length near the QD layer. As recombination-enhanced dislocation motion (REDM) processes are common failure mechanisms in semiconductor lasers, this is promising for laser reliability.

To investigate the efficacy of misfit trapping layers in full lasers, we fabricated InAs DWELL ridge structures on (001) Si with trapping layers in the epitaxial stack, shown schematically in Fig. 3a, alongside a baseline sample with no trapping layers, both grown from the same $3\times10^7$ cm$^{-2}$ TDD buffer (see references for buffer[11] and full laser[32] growth details). All lasers were fabricated together into 3-μm wide, 1500-μm long, cleaved-facet, deeply etched ridge structures. Unlike with the model structures, the GaAs/AlGaAs layers above the active region here are sufficiently thick to relax during cooldown, enabling MD formation at both the uppermost and lowermost DWELLs, as seen in Ref. 20.[20] To trap defects from both sides, we insert two sub-critical thickness 7-nm trapping layers 80 nm above and below the active region, composed of In$_{0.15}$Ga$_{0.85}$As and In$_{0.15}$Al$_{0.85}$As, respectively. These dissimilar alloys minimize electrical barriers due to band misalignment, but we expect them to have near-identical trapping ability. Since the covalent single-bond radii of aluminum (126 pm) and gallium (124 pm) are nearly identical, an equivalent indium alloying fraction should harden both layers similarly; note that these similar covalent radii also explain why AlGaAs alloys experience no hardening. Fig. 3b and 3c show the effect of trapping layers on MD formation via bright-field (BF) on-zone ([100]) cross-sectional STEM. All STEM images were acquired using a Thermo Fisher Talos 200X G2 TEM/STEM (200 kV) with a standard BF STEM circular detector and beam convergence angle of 10.5 mrad. The sample lift-out geometry, oriented at 45° to the orthogonal MD arrays (Fig. 3b inset), ensures that all MDs appear as equal-length horizontal lines. As shown in both low (Fig. 3b) and high (Fig. 3c) magnification images, MDs (black arrows) are displaced from the active region to the upper and lower trapping layers. Figure S2 provides additional evidence of misfit trapping in a full laser.

We illustrate the differences between the trapping layer and baseline structures using strain-contrast electron tomography generated from BF plan-view (PV)-STEM images taken across ~60° tilt range. Tomography is traditionally performed by tilting along a single axis, but here we followed the **g** = 220 Kikuchi band using a double-tilt holder to maximize strain contrast and used the BF detector as a virtual aperture. A sample PV-STEM image for the baseline (Fig. 3d) shows a MD amid a field of QDs. The tomographic reconstruction (Fig. 3e), created with Tomviz

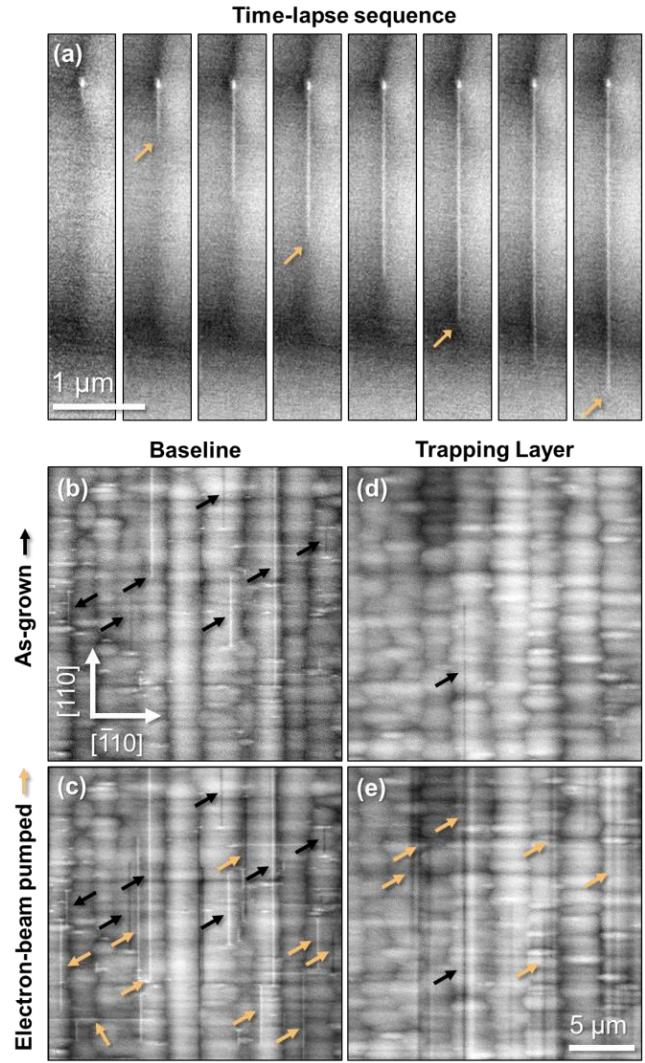

**Fig. 2.** **(a-e)** Electron-channeling contrast imaging (ECCI) of the baseline and trapping layer model structures. **(a)** ECCI time-lapse sequence (~550 s image interval) showing growth of a misfit dislocation (MD) (orange arrows) from a stationary, pinned threading dislocation (TD) segment in the baseline structure. **(b-c)** Corresponding ECCI of the baseline structure **(b)** before and **(c)** after electron-beam illumination. Black arrows indicate as-grown MD positions; orange arrows indicate MD growth from electron-beam pumping. **(d-e)** ECCI of the trapping layer structure **(d)** before and **(e)** after electron-beam illumination. Compared to the sharp line contrast of MDs in (b-c), the diffuse line contrast in (e) is due to MDs lying deeper in the structure.

(https://tomviz.org), resolves the five QD layers and shows that this MD lies at the uppermost QD layer. In a trapping layer laser, Fig. 3f and 3g show a PV-STEM image and a tomographic reconstruction, respectively, of a MD and a terminating TD segment. Although strain-contrast tomography cannot resolve the trapping layer itself, the MD clearly lies away from the DWELL at the trapping layer's height. The TD forms no additional MD segments as it travels downward through the QD layers. Fig. 3h also shows a PV-STEM image of a dislocation in a trapping layer laser, but here, there is a short, angled section along the MD, indicating a change in height. The tomographic reconstruction (Fig. 3i) confirms that the MDs lie at the trapping layer and the uppermost QD layer. This may be because TDs, normally pinned by the trapping layer, can become unpinned during cooldown and glide before becoming pinned again, explored further in Fig. S1. Unfortunately, this causes a MD to form at the outermost QD layer. Nevertheless, trapping layers successfully displace most MD length from the QDs as confirmed with PV-STEM (not shown).

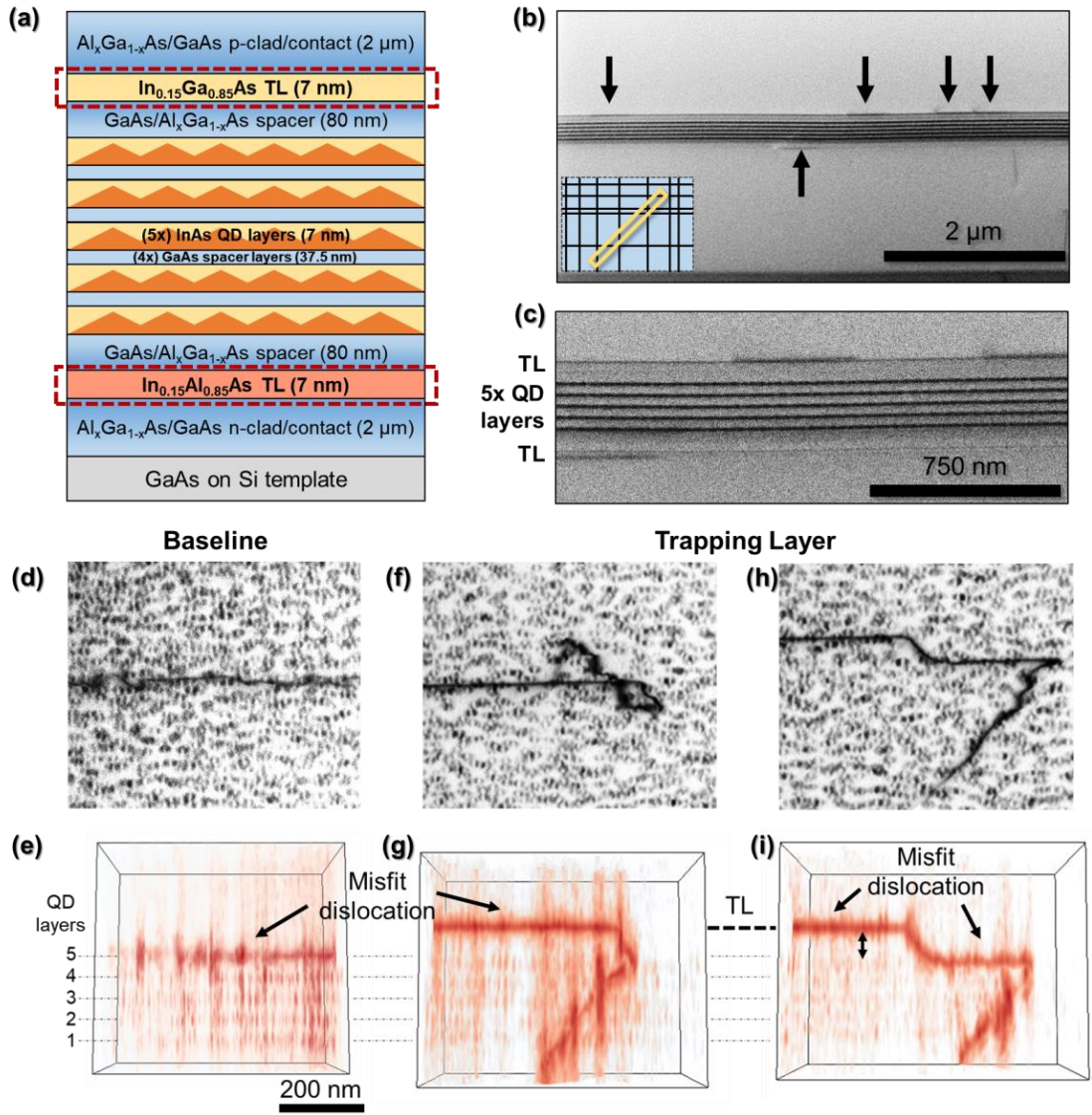

**Fig. 3.** (a) Schematic of a quantum dot (QD) laser with trapping layers (red boxes) above and below the QD layers. Baseline samples are equivalent but lack trapping layers. (b) Cross-sectional bright-field (BF) STEM ([100] zone) of a trapping-layer laser. Inset shows orientation of foil relative to misfit dislocations (MDs). Arrows mark MD segments at the trapping layers. (c) High-magnification image of (b). (d-e) Baseline laser: (d) BF plan-view (PV)-STEM image (**g = 220**) showing a MD among QDs. (e) Cross-sectional tomographic reconstruction showing the MD at the fifth QD layer. (f-i) Trapping layer laser: (f) BF PV-STEM showing a MD terminating in a threading dislocation (TD). (g) Reconstruction shows the MD lying at the trapping layer. (h) MDs at two heights with a TD end. (i) Reconstruction reveals a short MD at the top QD layer with the rest lying at the trapping layer.

Room temperature photoluminescence spectroscopy and continuous-wave (CW) light output-current-voltage (LIV) curves of a representative high performing device from each design are shown in Fig. 4a and 4b, respectively. Introducing trapping layers increases photoluminescence intensity by approximately 2× compared to baseline (Fig. 4a). This agrees with the marked improvements in threshold current, slope efficiency, and output power shown in the representative LIV curves (F ig. 4b). Histograms comparing the structures along these same metrics (Fig. 4c-4e) further support these performance improvements. The trapping layer design shows a 2× reduction in median threshold current from baseline. The lowest measured threshold current (16 mA) represents a 40% decrease from baseline minimum. This is also 20% below identically designed state-of-the-art lasers on Si,[32] even with a 4× higher TDD here. We additionally observe an impressive 60% increase in median slope efficiency and a 3.4× increase in median peak single-facet output powers for trapping layer lasers. Finally, the median electrically dissipated power at rollover for trapping layer lasers (0.85 W) is approximately twice that of baseline (0.46 W) (not shown). This indicates—assuming comparable thermal impedances—that the inclusion of trapping layers has increases the lasers' optical amplification (gain). We cannot determine whether trapping layers adversely impact electrical transport in these lasers due to large variability in the series resistances across both sets of devices. But as higher-than-usual specific contact resistances across all devices (p: $2.3 \times 10^{-5}$ $\Omega \cdot cm^2$, n: $5.5 \times 10^{-5}$ $\Omega \cdot cm^2$) represent a limiting factor on output power, we anticipate processing modifications will further improve device performance.

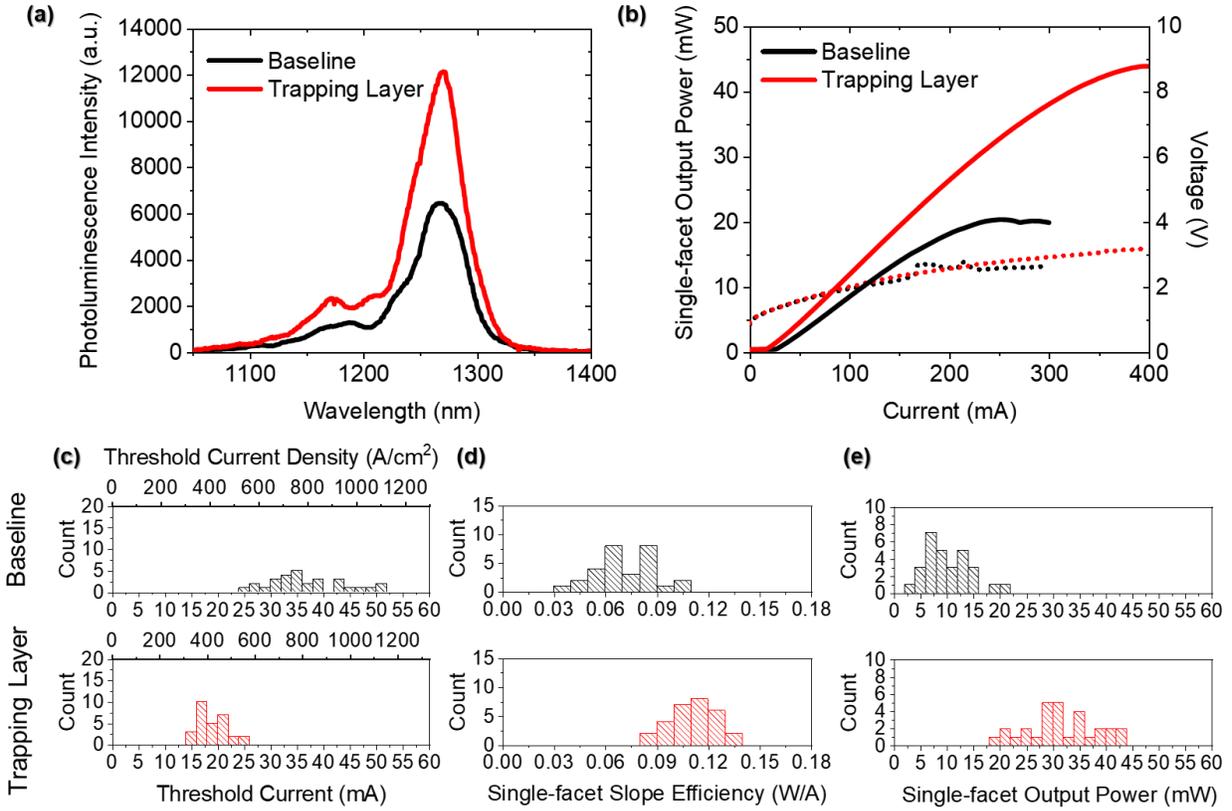

**Fig. 4. (a-e)** Comparison of baseline (black) and trapping layer (red) lasers. **(a)** Photoluminescence intensity comparison of trapping layer and baseline lasers. **(b)** Single-facet output power (mW) (solid) and voltage (V) (dashed) as a function of current (mA). A lower threshold current and higher slope efficiency and peak output power are observed in the trapping layer laser compared to baseline. Current-voltage (IV) curves are comparable for both designs. **(c-e)** Histograms showing performance improvements of trapping layer devices along key performance metrics: **(c)** threshold current (mA), **(d)** slope efficiency (W/A), and **(e)** output power (mW).

The relative performance improvements reported here—achieved simply by displacing existing MDs—are comparable to previous gains achieved by reducing TDD by an order-of-magnitude ($7 \times 10^7$ cm$^{-2}$ to $7 \times 10^6$ cm$^{-2}$).[33] As device thicknesses are critical for many applications, it is highly advantageous that these performance gains made using thin misfit trapping layers compare favorably to those achieved using hundreds of nanometers of traditional TD filters. This same single order-of-magnitude reduction in TDD also resulted in a nearly four order-of-magnitude increase in device lifetimes.[33] In these low TDD and low strain systems, the total active-region MD line length is determined by TDD and glide kinetics, so this dramatic increase in lifetime is likely explained in part by an unseen reduction in total MD line length. All dislocation line length, whether MD or TD, within the active region degrades laser performance and lifetime; the inclusion and optimization of trapping layers thus complements important, ongoing TDD reduction efforts.[34] In future work, we will determine whether eliminating MDs enables epitaxially integrated InAs QD lasers to finally meet commercial lifetime requirements at 60 °C operating temperature.

In summary, we have proposed a mechanism that describes how TDs to give rise to highly damaging MDs that form during post-growth cooldown in certain epitaxial III-V-on-silicon structures. We mitigate this by inserting thin alloy-hardened layers to pin TDs and displace MD formation away from the QDs, removing 95% of MD length in model structures. The trapping layers, placed both above and below the active region, represent a significant departure from traditional defect filtering: they displace, rather than remove, defects that form during cooldown, rather than during growth. For silicon photonics, this may finally clear the path to commercially viable, monolithically integrated, III-V-on-silicon photonic integrated circuits.

**SUPPLEMENTARY MATERIAL**

See supplementary material for (Fig. S1) schematic representations of (a) the stresses a dislocation experiences in our misfit trapping layer structures and the approximate stress landscapes in the (b) alloy-hardened $In_{0.15}Ga_{0.85}As$ misfit trapping layer and (c) the QD layer, (Fig. S2) two tilted cross-sectional bright field STEM images of threading dislocations that have formed trapped misfit segments in a trapping layer laser (**g** = **002**).

**ACKNOWLEDGEMENTS**

Sample growth was supported by ARPA-E, U.S. Department of Energy, under Award No. DE-AR00000843. This study is based upon work supported by the National Science Foundation Graduate Research Fellowship under Grant No. 1650114. Further support has been provided by the University of California, Santa Barbara Graduate Division through the Doctoral Scholars Program. K.M. acknowledges support from the California Nanosystems Institute SEED-TECH program. The research reported here made use of shared facilities of the UCSB MRSEC (NSF DMR 1720256), a member of the Materials Research Facilities Network. The authors are grateful to Arthur C. Gossard, Rushabh Shah, and Mario Dumont for the helpful discussions.

**AUTHOR CONTRIBUTIONS**

J.S., J.N., and E.T.H contributed equally to this work.

**DATA AVAILABLILITY**

The data that support the findings of this study are available from the corresponding author upon reasonable request.

# Supplementary Material

## Defect filtering for thermal expansion induced dislocations in III-V lasers on silicon

Jennifer Selvidge[1], Justin Norman[1], Eamonn T. Hughes[1], Chen Shang[1], Daehwan Jung[2], Aidan A. Taylor[1], MJ Kennedy[3], Robert Herrick[4], John E. Bowers[1,3], Kunal Mukherjee[1]

[1]Materials Department, University of California, Santa Barbara, CA 93106, USA
[2]Korea Institute of Science and Technology, Seoul 02792, South Korea
[3]Electrical and Computer Engineering Department, University of California, Santa Barbara, CA 93106, USA
[4]Intel Corporation, Santa Clara, CA 95054, USA

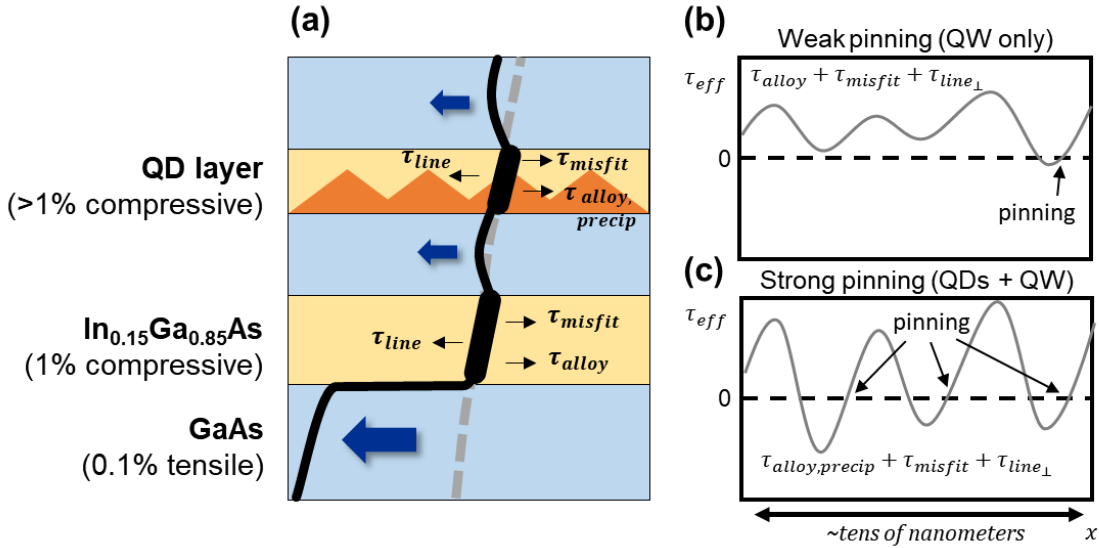

**Fig. S1.** (a) Schematic showing a dislocation traveling upwards through a GaAs-based film on Si, pinned in the trapping layer and the QD layer. $\tau_{line}$ represents the shear due to dislocation line tension; $\tau_{misfit}$, the shear due to lattice mismatch between GaAs and the strained indium-alloyed layers; and $\tau_{alloy}$, a resistive alloy hardening shear due to alloy compositional fluctuations. **(b-c)** Rough sketch of the effective stress landscape in **(b)** the $In_{0.15}Ga_{0.85}As$ trapping layer where pinning is relatively weak and **(c)** the QD layer, where the combination of QDs inside a QW results in strong pinning.

We analyze a simplified case of pinning (Fig. S1a) where a threading dislocation is completely mobile in GaAs and pinned in both the $In_{0.15}Ga_{0.85}As$ trapping layer and the QD layer above it. To glide in GaAs, the threading dislocation segment only needs to overcome the short-range, interatomic Peierls stress, $\tau_p$ (~4 GPa in GaAs[35]). This happens readily with relatively small resolved shear stresses either at elevated temperatures or through REDM processes,[36] which we exploit for ECCI in Fig. 2.

The stress states in the two indium-alloyed layers are more complex. We therefore employ the concept of an effective stress ($\tau_{eff}$), where, by our convention, the threading segments in these indium-alloyed layers can only glide leftward (along with the unpinned segment in the GaAs) if $\tau_{eff}$ is positive.[37–39] During cooldown, the sub-critical thickness $In_{0.15}Ga_{0.85}As$ and DWELL layers remain compressively strained. The threading segments in these layers experiences a shear stress ($\tau_{misfit}$) due to this strain, but since the layers are below critical thickness, $\tau_{misfit}$ must by definition be smaller than the line tension of the dislocation ($\tau_{line}$), or, more exactly, smaller than the maximum value of $\tau_{line}$ which is assumed when the threading segment forms a near-perpendicular kink ($\tau_{line_\perp}$). Without any additional resistive shear stresses, the shear from the dislocation line tension would normally drag these short threading segments along with it—no misfit segments would form.[39] Clearly, this is not the case for either the trapping layer or the QD layer.

---

[a] Author to whom correspondence should be addressed: kunalm@ucsb.edu

For pinning to occur as we observe, there must be an additional stress that adds to $\tau_{misfit}$ to at least match the magnitude of $\tau_{line_\perp}$. The source of this additional stress in the trapping layer is alloy hardening ($\tau_{alloy}$), which arises in certain semiconductor alloys due to natural compositional variations that generate in-layer stress fluctuations. Thus, the effective stress state in the trapping layer resembles that shown in Fig. S1b. This type of hardening effect has been observed in SiGe,[29] GaAsP,[30] and low-indium InAlGaAs alloys,[26] in agreement with our results. In our case, the alloy hardening effect results from the 21% volume difference between the $InAs_4$ and $GaAs_4$ tetrahedra[28]. Note that an alloy like AlGaAs, where the $AlAs_4$ and $GaAs_4$ tetrahedra are of near-identical size, should have no alloy hardening effect. In Fig. S1b, if the long-range resistive stress field, $\tau_{alloy}$, is large enough such that at some point $\tau_{eff} = \tau_{line_\perp} - (\tau_{misfit} + \tau_{alloy}) = 0$, then the threading dislocation segment is pinned in the trapping layer, and a trapped misfit dislocation segment will form. Note that the misfit segment cannot simply glide upward through the trapping layer due to the repulsive compressive strain in that layer.

The magnitude of the stress field in the QD layer is substantially larger than in the trapping layer, as shown in Fig. S1c. Alloy hardening in the QW once again provides a resistive shear, but, as Beanland et al. have shown, precipitate-like QDs also provide their own resistive shear, pinning threading segments so effectively that they nearly triple the critical thickness for dislocation glide compared to a QW.[27] These effects agree with metallurgical research showing that mechanical properties of both elemental metals and alloys become increasingly temperature independent, or athermal, with increasing temperature. This is because long-range fluctuating stress fields, generated both by compositional fluctuations and structural features such as precipitates and line defects, are no easier to surmount at high temperatures than at low ones.[40,41]

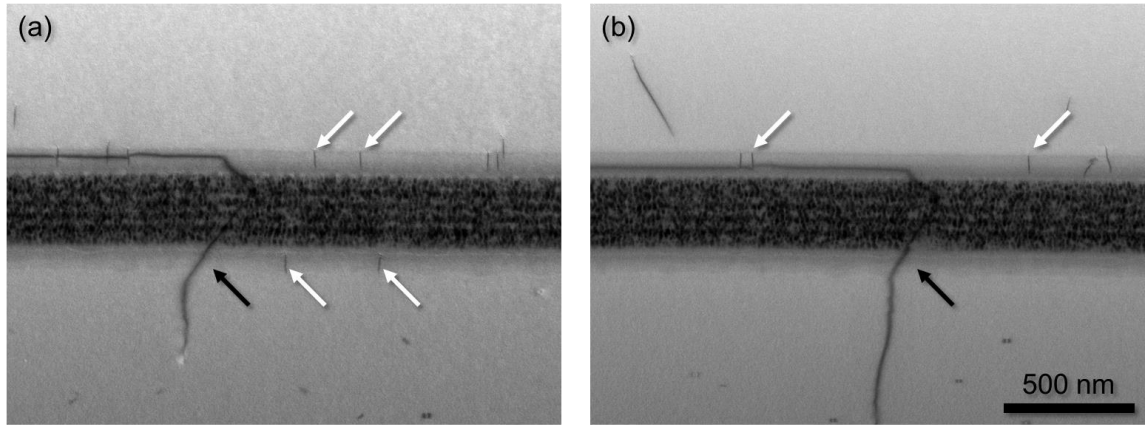

**Fig. S2.** Both (a) and (b) show cross sectional scanning transmission electron micrographs of both misfit and threading dislocations in a trapping layer laser. Samples were lifted out along [110] direction and imaged at a tilt (**g = 002**) resulting in certain misfits running parallel to the length of the foil and others running perpendicular to it (marked with white). Due to the tilt, the perpendicular misfits appear as vertical lines and the spacer layers between the quantum dots (QDs) disappear among the QD strain contrast. Critically, the misfit dislocations are clearly at different heights than the QDs. Threading dislocations, marked with black arrows, give rise to the misfit dislocations, as described.

Fig. S2 shows two images where a threading dislocation (marked with black arrow) passes through the QD layers and gives rise to a misfit segment at the top trapping layer (analogous to Fig. 3f-3i). Due to the tilt, these two misfit segments appear to lie at different heights, but they actually both lie at the trapping layer. From this, we can infer that the misfit segments lie at different depths from the face of the foil. In both images, we can additionally see misfit dislocations lying in the direction of the foil thickness (marked with white arrows). From the length of these misfits and tilt angles, we can determine the foil thickness. It is worth noting, in both images, that QDs adjacent to the threading dislocation appear slightly different from the others both in density and in appearance. This confirms that the threading dislocation has not moved from its growth position in the QD layers. The consistency between these images and Fig. 3b and 3f-3i provides strong evidence for the success of both the upper and lower trapping layer.